\documentclass[conference]{IEEEtran}
\IEEEoverridecommandlockouts
\usepackage{amsmath,amssymb,amsfonts}
\usepackage{algorithmic}
\usepackage{graphicx}
\usepackage{textcomp}
\usepackage{lipsum}
\usepackage{flushend}
\usepackage{tabularx}
\usepackage{multirow}
\usepackage{booktabs}
\usepackage[breaklinks]{hyperref}
\usepackage[linesnumbered,ruled,lined,commentsnumbered]{algorithm2e}
\usepackage{color,soul}
\usepackage[T1]{fontenc}
\usepackage[utf8]{inputenc}
\usepackage{authblk}
\usepackage[square, comma, sort&compress,numbers]{natbib}
\usepackage[dvipsnames]{xcolor}
\usepackage{graphicx}

\def\BibTeX{{\rm B\kern-.05em{\sc i\kern-.025em b}\kern-.08em
    T\kern-.1667em\lower.7ex\hbox{E}\kern-.125emX}}
\begin{document}

\title{RankDVQA-mini: Knowledge Distillation-Driven Deep Video Quality Assessment\thanks{Research reported in this paper was supported by an Amazon Research Award, Fall 2022 CFP. Any opinions, findings, and conclusions or recommendations expressed in this material are those of the author(s) and do not reflect the views of Amazon. We also appreciate the funding from the China Scholarship Council, University of Bristol, and the UKRI MyWorld Strength in Places Programme (SIPF00006/1).}}

\author[1]{Chen Feng}
\author[1]{Duolikun Danier}
\author[1]{Haoran Wang}
\author[1]{Fan Zhang}
\author[2]{Benoit Vallade}
\author[2]{Alex Mackin}
\author[1]{David Bull}
\affil[1]{\textit{Visual Information Laboratory, University of Bristol, Bristol, BS1 5DD, United Kingdom}}
\affil[1]{\textit {\{chen.feng, duolikun.danier, yp22378, fan.zhang, dave.bull\}@bristol.ac.uk}}
\affil[2]{\textit{Amazon Prime Video, 1 Principal Place, Worship Street, London, EC2A 2FA, United Kingdom}}
\affil[2]{\textit {\{valladeb, acmackin\}@amazon.co.uk}}

\maketitle

\begin{abstract}
Deep learning-based video quality assessment (deep VQA) has demonstrated significant potential in surpassing conventional metrics, with promising improvements in terms of correlation with human perception. However, the practical deployment of such deep VQA models is often limited due to their high computational complexity and large memory requirements. To address this issue, we aim to significantly reduce the model size and runtime of one of the state-of-the-art deep VQA methods, RankDVQA, by employing a two-phase workflow that integrates pruning-driven model compression with multi-level knowledge distillation. The resulting lightweight full reference quality metric, RankDVQA-mini, requires less than 10\% of the model parameters compared to its full version (14\% in terms of FLOPs), while still retaining a quality prediction performance that is superior to most existing deep VQA methods. The source code of the RankDVQA-mini has been released at \url{https://chenfeng-bristol.github.io/RankDVQA-mini/} for public evaluation.
\end{abstract}

\begin{IEEEkeywords}
Video quality assessment, deep learning, model compression, knowledge distillation, RankDVQA-mini
\end{IEEEkeywords}

\section{Introduction}

Objective video quality assessment plays an essential role in many video processing applications \cite{bull2021intelligent}. It can, for example, be used for comparing the performance of compression algorithms, or within these algorithms to guide model optimisation (e.g., in rate-quality optimisation or as a loss function for training learning-based methods) \cite{ndjiki2012perception,ma2020cvegan}. Over the last two decades, VQA methods have experienced significant advances, from the conventional quality metrics based on classic signal measures to more recent contributions optimised using deep learning techniques.

The most commonly used conventional quality metrics\footnote{In this work, we solely focus on full reference scenarios where the reference content of the distorted video is available.} are PSNR and SSIM \cite{ssim}, which measure pixel-wise distortions and the similarity between the test and reference content, respectively. To improve their correlation with visual perception, many perceptual-inspired objective quality metrics have been developed, including SSIM variants \cite{c:mssim,ssimplus,vssim}, VQM \cite{VQM}, MOVIE \cite{MOVIE}, PVM \cite{PPVM} and MAD \cite{MAD}. Enhanced methods exist \cite{lin2014fusion,zhang2021enhancing} that integrate these perceptual quality metrics together with video features into a regression-based framework to further improve the overall correlation with subjective opinions. One important metric in this class is VMAF \cite{w:VMAF}. VMAF has been widely adopted for evaluating the performance of video compression and processing algorithms due to its excellent and consistent performance.

\begin{figure}[t]
    \centering
    \includegraphics[width=\linewidth]{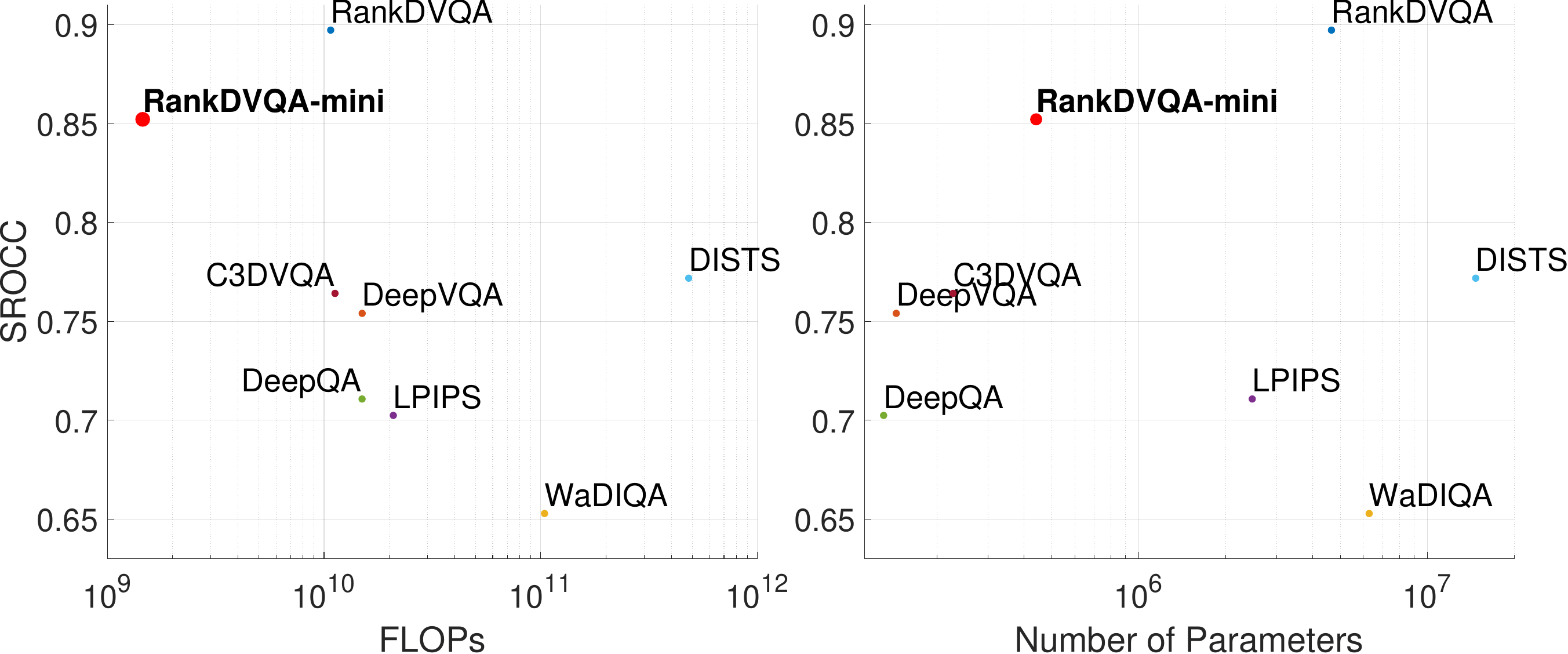}
    \caption{Complexity (in terms of either FLOPs or model size) versus correlation performance (in SROCC) plot for benchmarked deep VQA methods. FLOPs are calculated for inference on a 256$\times$256 sequence of 12 frames.}
    \label{fig:star}
    \vspace{-10pt}
\end{figure}

More recently, VQA metrics have been further enhanced through deep learning techniques. Various metrics have been reported, which have been built using Convolutional Neural Networks (CNNs); notable examples include DeepVQA \cite{Kim2018DeepVQ}, C3DVQA \cite{C3DVQA}, LPIPS \cite{Lpips} and DISTS \cite{DISTS}. Although these methods have been reported to offer promising results compared to conventional and regression-based VQA methods, they are constrained by the lack of reliable large and diverse training databases (the model is typically trained on a small video quality dataset with limited subjective ground truth labels) and do not show consistent generalisation performance. To address this issue, RankDVQA \cite{feng2022rankdvqa} was proposed that employs a ranking-inspired training methodology. This supports the use of a large scale training database for model optimisation, and achieves the state-of-the-art generalisation performance compared to other existing methods. 


Although deep VQA methods offer the potential to outperform conventional and regression-based quality metrics, they are often associated with high computational complexity, which restricts their deployment in practical applications \cite{gou2021knowledge}. To address this challenge, network pruning \cite{reed1993pruning} and knowledge distillation \cite{wang2021knowledge, hinton2015distilling} techniques can be employed to prune a large network (denoted by ``teacher'') and transfer its knowledge to the pruned smaller model (the ``student''). Such techniques have been applied in numerous fields \cite{morris2023stmfnet, park2019relational, mishra2017apprentice}, demonstrating that a compact model can achieve comparable performance to its larger counterpart.

In this work, we apply these techniques to deep VQA, producing a lightweight network that achieves competitive performance without the need to incorporate additional modules. Specifically, we employ a two-stage pipeline to condense the RankDVQA \cite{feng2022rankdvqa} network. RankDVQA is selected due to its state-of-the-art performance and model generalisation capabilities. Inspired by previous work \cite{ding2021cdfi, morris2023stmfnet}, we first apply a sparsity-inducing pruning technique to substantially reduce the number of RankDVQA parameters - retaining only 10\% of the original model's parameters. The pruned model is then trained using a multi-level knowledge distillation strategy \cite{jin2023multi}, learning from the teacher model, a pre-trained original RankDVQA. As shown in Fig. \ref{fig:star}, the final compact model, RankDVQA-mini, retains 96\% of RankDVQA's performance in terms of SROCC, while removing  90.12\% of its parameters. Moreover, RankDVQA-mini reduces the Floating Point Operations (FLOPs) count of the original model by 86.42\% (from 10.731G to 1.457G).  We believe that this is the first time that model pruning and knowledge distillation have been used to optimise a deep VQA metric -- a step towards low-complexity deep VQA. 


\section{Proposed Method}
\label{sec:Proposed method}


RankDVQA has been selected as the anchor VQA model for complexity reduction in this work. RankDVQA \cite{feng2022rankdvqa} is one of the latest and best performing learning-based video quality metrics. It is based on a ranking-inspired training strategy that enables the development of large and reliable training databases without performing expensive subjective tests. It consistently achieves superior performance compared to other conventional, regression-based and deep VQA methods. 

The architecture of RankDVQA consists of two parts: the PQANet, which uses convolutional and SWIN transformer \cite{liu2021swin} layers for feature extraction and local quality prediction, and the STANet, which refines the assessment using adaptive spatio-temporal pooling. Since the model size of STANet (14.0K parameters) is much smaller than that of PQANet (4.59M), in this work we solely focus on reducing the complexity of the PQANet model.

\subsection{Pruning RankDVQA}

\textbf{Sparsity-inducing Optimisation.} Model pruning in the context of PQANet aims to simplify the network by inducing sparsity in parameters, thereby providing guidance for removing unnecessary model parameters while maintaining its performance. This is achieved by adding an L1 norm regularisation term to the training loss function, $\mathcal{L}_{prune}$, and applying the OBProx-SG optimiser \cite{chen2021orthant}:
\begin{equation}
\mathcal{L}_{prune} = \mathcal{L}_{s1} + \lambda \cdot ||\theta||_{1}. \label{eqn:lprune}
\end{equation}
Here, $\mathcal{L}_{s1}$ represents the original binary cross entropy loss for training the PQANet \cite{feng2022rankdvqa}. $ \lambda $ is a hyper-parameter that controls the sparsity level (its empirical value equals 0.1). $\theta$ refers to the parameters of the PQANet model \cite{chen2021orthant}.

The optimisation process here is expected to identify any relatively unimportant parameters. After around 30 epochs, the sparse model retains those parameters that make a relatively significant contribution to overall model performance. The number of non-zero parameters reduces from around 4.59 million to 0.44 million. This compact model serves as a starting point for subsequent model compression. 

\textbf{Model Pruning}. The sparse model obtained above is an important indicator for model pruning. We define the density of each layer, $\mathcal{D}_L$, as the proportion of its non-zero parameters:
\begin{equation}
\mathcal{D}_L = \frac{\text{number of nonzero parameters in } L}{\text{total parameters of } L},
\end{equation}
where $L$ denotes a specific layer with parameters. Consequently, the contribution of each layer to the model performance can be identified. In the pruning stage, the density value is employed as the compression ratio of each layer.

To prune the model, the redundant channels in the original model are removed, starting from the last layer and proceeding backwards. The number of input channels $C_{in, L}$ for any given layer, $L$, is reduced to $\mathcal{D}_L\cdot C_{in,L}$, with $\mathcal{D}_L$ symbolising the compression ratio equivalent to the layer's density. Adjacent layers are then adjusted in tandem to maintain the integrity of the network. As a result, the number of output channels in these layers is decreased proportionally. The final compact PQANet model contains only 9.58\% of the original model weights due to the significant reduction of redundant channels. 


\subsection{Multi-level Knowledge Distillation}

\begin{figure*}[htbp]
    \centering
    \includegraphics[width=0.9\linewidth]{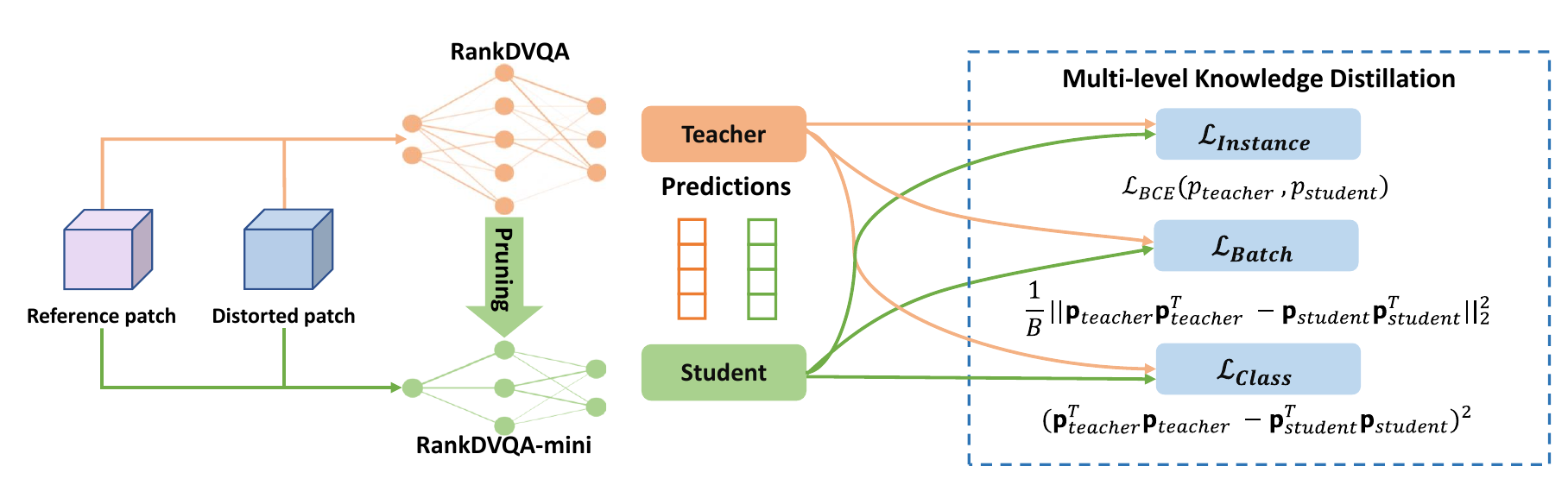}
    \vspace{-5pt}
    \caption{The framework of the RankDVQA-mini with the multi-level knowledge distillation, after the model pruning and compression, obtaining the RankDVQA (teacher) and RankDVQA-mini (student) predictions. The predictions are matched respectively through multi-level alignment, which consists of instance-level, batch-level, and class-level alignments. We take batch size B = 4 as an example to demonstrate this approach.}
    \label{fig:Multilevel}
\end{figure*}

After model pruning, knowledge distillation is employed to enhance the performance of the compact PQANet model. Our approach differs from the traditional knowledge distillation framework \cite{morris2023stmfnet, wang2021knowledge, park2019relational, hinton2015distilling}, which only focuses on the difference between the output of student and teacher models at the instance level. Inspired by recent work \cite{jin2023multi}, we adopt a multi-level logit knowledge distillation strategy, which extends this process to two additional levels: batch and class, to enhance the learning efficiency of the student model from the teacher. Fig. \ref{fig:Multilevel} shows the workflow of this approach.

\textbf{Instance-level Alignment}. This level inherits the conventional knowledge distillation method \cite{morris2023stmfnet, park2019relational, hinton2015distilling}, which minimises the divergence between the outputs of the teacher and student models at the instance level. During optimisation, the student model is trained to decrease the difference in prediction from the teacher model for each instance. Specifically, PQANet takes as input (in the training process - details can be found in \cite{feng2022rankdvqa}) two distorted sequences $\mathbf{D}_1, \mathbf{D}_2$ and their respective references $\mathbf{R}_1, \mathbf{R}_2$, and outputs the probability of $\mathbf{D}_1$ having higher quality than $\mathbf{D}_2$. Let $p_{teacher}$ denote the output of the teacher network (the original PQANet) on one such training instance, and let $p_{student}$ be the output of the student network (the pruned PQANet) for the same instance. The instance level loss function reads: 
\begin{equation}
    \mathcal{L}_{\mathrm{Instance}} = \mathcal{L}_\mathrm{BCE}(p_{teacher},p_{student}).
\end{equation}
Here the Binary Cross Entropy (BCE) loss measures the divergence between $P_{teacher}$ and $P_{student}$, 
\begin{multline}
     \mathcal{L}_{\mathrm{BCE}} =  - \big(p_{teacher}\log(p_{student}) \\+ (1-p_{teacher})\log(1-p_{student})\big).
\end{multline}

\textbf{Batch-level Alignment}. At the batch level, we aim to train the student to mimic the inner-instance correlation (within a training batch) predicted by the teacher. Specifically, let $\mathbf{p}_{teacher}, \mathbf{p}_{student} \in(0,1)^{B\times 1}$ denote the outputs of the teacher and the student on a batch of $B$ training instances, the batch-level distillation loss is defined as
\begin{equation}
\mathcal{L}_{\mathrm{Batch}} = \frac{1}{B}||\mathbf{p}_{teacher}\mathbf{p}_{teacher}^T -\mathbf{p}_{student}\mathbf{p}_{student}^T||^2_2,
\end{equation}
where the Gram matrix $\mathbf{p}\mathbf{p}^T\in (0,1)^{B \times B}$ models the inner-instance relationships.

\textbf{Class-level Alignment}. The class-level alignment in \cite{jin2023multi}, matches the category correlation modelled by the student and the teacher, i.e. the relationship between different classes (e.g. the 1000 classes in ImageNet~\cite{russakovsky2015imagenet}). In our case, the classification is binary, and the loss function is  written as:
\begin{equation}
\mathcal{L}_{\mathrm{Class}} = (\mathbf{p}_{teacher}^T\mathbf{p}_{teacher} -\mathbf{p}_{student}^T\mathbf{p}_{student})^2,
\end{equation}

\textbf{Multi-level Loss}. Three levels of loss functions will be combined to act as the knowledge distillation term in the training loss function: 
\begin{equation}
    \mathcal{L}_{\mathrm{Multi-level}} = \mathcal{L}_{\mathrm{Instance}} + \mathcal{L}_{\mathrm{Batch}} + L_{\mathrm{Class}},
\end{equation}
\begin{equation}
\mathcal{L}_\mathrm{total} = \mathcal{L}_{\mathrm{Multi-level}} + \alpha \mathcal{L}_{\mathrm{s1}},
\end{equation}
where $\mathcal{L}_{\mathrm{s1}}$ denotes the original loss function~\cite{feng2022rankdvqa} for the PQANet as mentioned in Equation~(\ref{eqn:lprune}), and $\alpha$ is a hyperparameter to allocate weight to the distillation loss term (set as 0.1). Our method combines three losses at different levels, guiding the student model to mimic the teacher model's behaviour across instance-level, batch-level and class-level predictions.

\begin{table*}[t]
	\scriptsize
	\centering
	\caption{Performance and complexity figures of the proposed method and other benchmark approaches on eight test databases. The values in each cell x(y) correspond to the SROCC value (x) and F-test result (y) at 95\% confidence interval. y=1 suggests that the metric is superior to RankDVQA-mini (y=-1 if the opposite is true), while y=0 indicates that there is no significant difference.}
 \setlength{\tabcolsep}{1.75mm}{
	\begin{tabularx}{\linewidth}
		{r||>{\centering\arraybackslash}
		l>{\centering\arraybackslash}
		l>{\centering\arraybackslash}
		l>{\centering\arraybackslash}
		l>{\centering\arraybackslash}
		l>{\centering\arraybackslash}
		l>{\centering\arraybackslash}
		l>{\centering\arraybackslash}
		l|>{\centering\arraybackslash}
		l|>{\centering\arraybackslash}
		c|>{\centering\arraybackslash}
		c}
		\toprule
		\centering
        \setlength{\tabcolsep}{12pt}
		SROCC(F-test) & NFLX & NFLX-P & BVI-HD & CC-HD & CC-HDDO & MCL-V & SHVC  & VQEGHD3 & \textbf{Overall} & FLOPs (G) & \#P (M)\\
		\midrule \midrule
        PSNR  & 0.6218 (-1) & 0.6596 (-1) & 0.6143 (-1) & 0.6166 (-1) & 0.7497 (-1) & 0.4640 (-1) & 0.7380 (-1)  & 0.7518 (-1) & 0.6520 & ---&--- \\\midrule
		SSIM \cite{ssim} & 0.5638 (-1) & 0.6054 (-1) & 0.5992 (-1) & 0.7194 (-1) & 0.8026 (-1) & 0.4018 (-1) & 0.5446 (-1)  & 0.7361 (-1) & 0.6216 & --- &---\\\midrule
		MS-SSIM \cite{c:mssim} & 0.7136 (-1) & 0.7394 (-1) & 0.7652 (0) & 0.7534 (-1) & 0.8321 (0) & 0.6306 (-1) & 0.8007 (0)  & 0.8457 (0) & 0.7601 & --- &---\\\midrule
	    WaDIQA \cite{8063957}  & 0.5713 (-1) & 0.6593 (-1) & 0.6646 (-1) & 0.6516 (-1) & 0.7041 (-1) & 0.6072 (-1) & 0.6731 (-1)  & 0.6910 (-1) & 0.6528& 104.04 & 6.287\\\midrule
		DeepQA \cite{Kim_2017_CVPR} & 0.7298 (-1) & 0.6995 (-1) & 0.7106 (-1) & 0.6202 (-1) & 0.6705 (-1) & 0.6832 (-1) & 0.7176 (-1)  & 0.7881 (-1) & 0.7024&14.976&0.131 \\\midrule
		LPIPS \cite{Lpips}&0.6793(-1) &	0.7859 (-1)	& 0.6670 (-1)& 	0.6838 (-1)&	0.7678 (-1)&	0.6579 (-1)& 0.6360 (-1)	&0.8075 (0) & 0.7107&20.857& 2.472 \\\midrule	
		DeepVQA \cite{Kim2018DeepVQ} & 0.7352 (-1) & 0.7609 (-1) & 0.7330 (-1) & 0.6924 (-1) & 0.8120 (0) & 0.6142 (-1) & 0.8041 (0)  & 0.7805 (-1) & 0.7540& 14.990 &0.144\\\midrule
        C3DVQA \cite{C3DVQA} & 0.7730 (-1) & 0.7714 (-1) & 0.7393 (-1) & 0.7203 (-1) & 0.8137 (0) & 0.7126 (0) & 0.8194 (0)  & 0.7329 (-1) & 0.7641&11.236& 0.227 \\\midrule
	    DISTS \cite{DISTS} & 0.7787 (-1) & 0.9325 (0) & 0.7030 (-1) & 0.6303 (-1) & 0.7442 (-1) & 0.7792 (0) & 0.7813 (0) & 0.8254 (0) &  0.7718   & 481.07&14.715\\\midrule
        ST-GREED \cite{STGREED}& 0.7470 (-1) & 0.7445 (-1) & 0.7769 (0) & 0.7738 (0) & 0.8259 (0) & 0.7226 (0) & 0.7946 (0)  & 0.8079 (0) & 0.7842 &--- &---\\\midrule
		VMAF 0.6.1 \cite{w:VMAF}   &  0.9254 (0)   & 0.9104 (0)  &  0.7962 (0)&  0.8723 (0)&  0.8783 (0)&  0.7766 (0)&     0.9114 (0)   & 0.8442 (0) &  0.8644& --- &---\\\midrule
        FR-RankDVQA \cite{feng2022rankdvqa}   &  0.9393 (0)  &  0.9184 (0) &  0.8659 (0)  &  0.8991 (0) &  0.9037 (0)   &  0.8391 (0)  &  0.9142 (0)&  0.8979 (0) &  0.8972&10.731&4.608\\\midrule
	\textbf{RankDVQA-mini}    &  \textbf{0.8846}  & \textbf{0.8748}  &  \textbf{0.8135}  &  \textbf{0.8479}  &  \textbf{0.8890}   & \textbf{0.7592}   &  \textbf{0.8819} &  \textbf{0.8661} &  \textbf{0.8521}&\textbf{1.457}&\textbf{0.455}\\
		\bottomrule
        \end{tabularx}}
	\label{tab:results}
\end{table*}

\section{Results and Discussion}
\label{sec:results}


The compact version of PQANet was trained for 30 epochs on the same training set as the original model \cite{feng2022rankdvqa}, which consists of approximately 20K patch pairs from the CVPR 2022 CLIC video compression challenge \cite{clic} and BVI-DVC \cite{ma2021bvi} datasets. The STANet used for spatio-temporal pooling in the second stage remains the same, and has been retrained based on the output of PQANet, also with the same training databases as in \cite{feng2022rankdvqa}. We use AdaMAX optimisation \cite{kingma2014adam} with hyper-parameters $\beta1$=0.9 and $\beta2$=0.999 in the training process. Training and evaluation were executed on the compute cluster \cite{BC4} at the University of Bristol (GPU nodes with 2.4GHz Intel CPUs and two NVIDIA P100 graphic cards). 

To evaluate model generalisation performance, we followed the same experiment setup as in \cite{feng2022rankdvqa}, using eight different HD VQA datasets for performance benchmarking: NFLX \cite{w:VMAF}, NFLX-P \cite{w:VMAF}, BVI-HD \cite{j:Zhang7},  CC-HD \cite{katsenou2022bvi}, CC-HDDO \cite{c:Zhang24}, MCL-V \cite{j:Lin4}, SHVC \cite{r:JCTVCW0095}, VQEGHD3 \cite{r:vqegHD}. These databases contain various distortion types produced by spatial resolution adaptation and video compression.


To benchmark the performance of RankDVQA-mini we compared its correlation performance with eleven full reference quality assessment methods, including three conventional quality metrics: PSNR, SSIM \cite{ssim}, MS-SSIM \cite{c:mssim}\footnote{It is noted that these image quality metrics are calculated based on luma components only.}; and seven deep quality assessment methods\footnote{The selection of deep VQA methods is based on the performance reported in their original publications and on the availability of their pre-trained models.}: WaDIQA \cite{8063957}, DeepQA \cite{Kim_2017_CVPR}, DeepVQA \cite{Kim2018DeepVQ}, C3DVQA \cite{C3DVQA}, DISTS \cite{DISTS}, LPIPS \cite{Lpips}, RankDVQA \cite{feng2022rankdvqa}, and two regression-based VQA approach, ST-GREED \cite{STGREED} and VMAF \cite{w:VMAF}. 

To assess the correlation performance of these VQA methods with subjective ground truth, the Spearman Rank Order Correlation Coefficient (SROCC) was calculated, for each database, between predicted quality indices and subjective scores. Additionally, to test the significance of performance difference, an F-test was performed between the proposed method, RankDVQA-Mini, and other tested metrics based on the residuals between the predicted quality indices (after a non-linear regression) and the subjective ground truth \cite{PPVM,j:LIVE}.

Table \ref{tab:results} summarises the quantitative results of all tested VQA methods in terms of SROCC values, F-test results and complexity figures (number of model parameters and Floating Point Operations (FLOPs)). Note that the model size figures presented for RankDVQA and RankDVQA-mini are for both PQANet and the STANet. It can be observed that, with only 9.87\% of the parameters and 13.57\% of FLOPs compared to the original RankDVQA, RankDVQA-mini still achieves competitive correlation performance, outperforming all other deep VQA methods, including DISTS, C3DVQA and LPIPS, and conventional quality metrics: PSNR, SSIM and MS-SSIM. It is noted that RankDVQA-mini does not outperform VMAF (although the overall SROCC is competitive). However, this is the first step towards reducing the complexity of deep VQA metrics (which are the state of the art in terms of performance) for their practical use and we show a promising trade-off between complexity and performance. It is our hope to inspire further work on developing compact and efficient deep VQA models that surpass VMAF. Finally, according to the F-test, RankDVQA-mini shows significant advantage over most compared methods on various test sets, and its differences from VMAF and the original RankDVQA are insignificant.

Figure \ref{fig:star} provides a more intuitive comparison between RankDVQA-mini and other deep VQA methods in terms of performance and complexity. It can be observed that the proposed method achieves an excellent trade off between correlation performance and complexity (model size and FLOPs) - it requires a similar level (in the same order of magnitude) of model size and FLOPs as DeepVQA, C3DVQA and DeepQA, but achieves evident performance improvement (confirmed by the F-test results in TABLE \ref{tab:results}).

\section{Conclusion}
\label{sec:conclusion}

In this work, we present a new lightweight and effective deep video quality assessment method, RankDVQA-mini, by applying a two phase complexity reduction workflow to the state-of-the-art deep quality metric, RankDVQA. The resulting compact model retains the superior performance of its original counterpart, but with a reduction of 90\% in terms of model parameters and 14\% of FLOPs. Future work should focus on further runtime reductions and more effective knowledge distillation to improve model performance.

\small
\bibliographystyle{IEEEtran}
\bibliography{IEEEabrv, egbib}
\end{document}